\documentclass[11pt,letterpaper]{article}
\usepackage{arxiv}

\usepackage[english]{babel}     
\addto\captionsenglish{%
}
\usepackage{amsmath}            
\usepackage{epstopdf}           
\usepackage{flushend}           
\usepackage{hyperref}           
\usepackage{graphicx}
\usepackage{tabularx,multirow,booktabs,blindtext}
\usepackage{caption}
\usepackage{rotating}
\usepackage{float}
\usepackage{xcolor}
\usepackage{textcomp}
\usepackage{booktabs}
\usepackage{amssymb}
\usepackage{subfigure}
\def\BibTeX{{\rm B\kern-.05em{\sc i\kern-.025em b}\kern-.08em
    T\kern-.1667em\lower.7ex\hbox{E}\kern-.125emX}}
    
\begin{document}



\title{iCardo: A Machine Learning based Smart Healthcare Framework for Cardiovascular Disease Prediction}

\author{
Nidhi Sinha \\
ECE Department, \\
Malaviya National Institute of Technology \\
Dept. of DSE, Konverge.AI\\
Jaipur, India \\
\texttt{2018rec9033@mnit.ac.in} \\
\And
Teena Jangid \\
ECE Department, \\
Malaviya National Institute of Technology \\
Jaipur, India \\
\texttt{tinajangir1996@gmail.com}\\
\And
Amit M. Joshi \\
ECE Department, \\
Malaviya National Institute of Technology \\
Jaipur, India \\
\texttt{amjoshi.ece@mnit.ac.in} \\
\And
Saraju P. Mohanty \\
CSE Department\\
University of North Texas, \\
USA \\
\texttt{saraju.mohanty@unt.edu} \\
}

\maketitle











\begin{abstract}
Cardiovsacular Disease is major threat to humans around the world due to difficulty of the diagnosis at early stage. 
Real-time access to medical records is helpful to improve the patient's health in critical conditions. The point-of-care services and medication have become simpler with the use of efficient consumer electronics devices in a smart healthcare system.
Cardiovascular disease is one of the critical illnesses which causes the heart failure and the early and prompt identification of CVD can lessen damage and prevent premature mortality. Machine learning has been used to predict cardiovascular disease (CVD) in the literature. The article explains choosing the best classifier model for the selected feature sets and the distinct feature sets selected using four feature selection models. The paper also compares seven different classifiers using each of the sixteen feature sets. Originally, the data had 56 attributes and 303 occurrences, of which 87 were in good health and the remainder had cardiovascular disease (CVD). Demographic characteristics, Symptom and Examination features, Electrocardiography (ECG) based features, and Laboratory and Echocardiography based features make up the four groups that comprise the data set's overall features. Least Absolute Shrinkage and Selection Operator (LASSO), Tree-based algorithms, Chi-Square and Recursive Feature Elimination (RFE) have all been used to choose the four distinct feature sets, each containing five, ten, fifteen, and twenty features, respectively.
Seven distinct classifiers have been trained and evaluated for each of the sixteen feature sets. To determine the most effective blend of feature set and model, a total of 112 models have been trained, tested, and their performance metrics have been compared. 
A Support Vector Machine (SVM) classifier with fifteen chosen features is shown to be the best in terms of overall accuracy. The healthcare data has been maintained in the cloud and would be accessible to patients, caretakers, and healthcare providers through integration with the Internet of Medical Things ({IoMT}) enabled smart healthcare. Subsequently, the most appropriate feature for CVD prediction is chosen by feature selection model that is later utilised to calibrate the system, and the proposed framework can be utilised to anticipate CVD.
\end{abstract}




\section{Introduction}
\label{intro}

The advancement of healthcare technology has allowed the development of effective Healthcare Cyber-Physical Systems (H-CPS) which integrates electronic health records (EHR), with artificial intelligence (AI) for smart healthcare management.  The data from the healthcare sensors have allowed monitoring of the patient's condition through the integration of the Internet of Medical Things (IoMT). In \cite{CPS}, the intelligent feedback system through wearable sensors in healthcare, allows efficient decision-making for the point of care service. Disease diagnosis through remote monitoring is one of the most important components of smart healthcare \cite{Srivastava2022-pb, DWIVEDI2022302}. Cardiovascular Disease (CVD) is the leading cause of premature death worldwide, and the early stage of diagnosis through smart healthcare would allow for reducing mortality.

\begin{figure}[ht]
\centering
    \includegraphics[width=15cm]{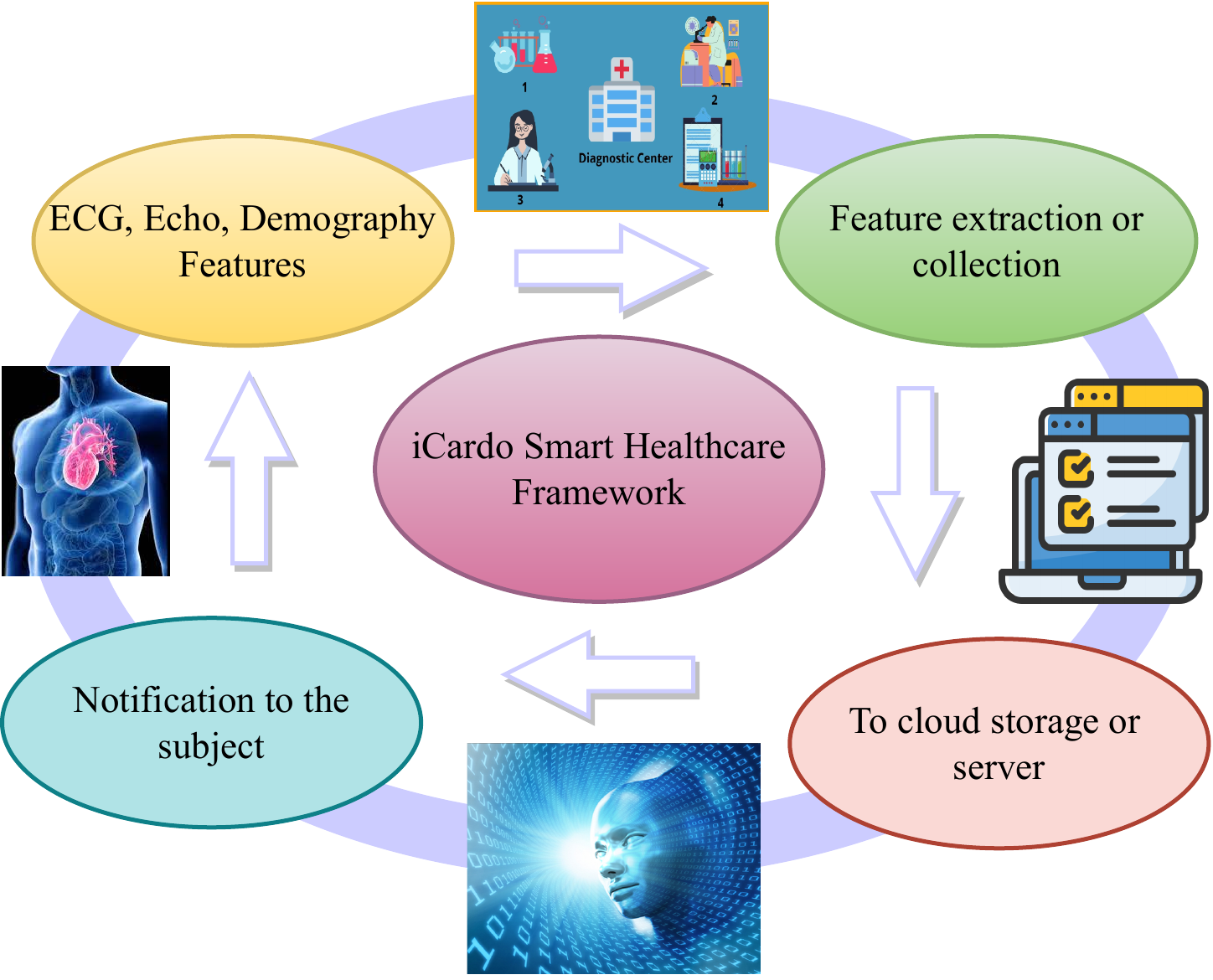}\\
    \caption{Smart healthcare framework for CVD prediction}
    \label{SHC}       
\end{figure}

According to the World Health Organization (WHO), Cardiovascular Diseases (CVDs) account for 33 \% of overall fatalities all over the world \cite{WHO2020}\cite{WHF2020}. 
The early detection of CVD would be helpful to have preventative measures for minimum premature mortality. In the past, various machine-learning models were attempted for automated detection of CVD \cite{hammad2020multitier,shomaji2019early,wu2020data, Nadakinamani2022}. The models would allow for providing an intelligent preventive healthcare mechanism. 
Such an automated system is capable of integration with
smart healthcare, which plays a vital role in the quality of life improvement \cite{iglu_TCE, sharma2022szhnn}. The portable medical equipment, such as patches, badges, rings, bracelets, and wrist devices, have offered simpler methods for everyday health monitoring \cite{smart_healthcare_review, Mywear, iKardo}. A novel non-invasive device to measure the amount of glucose and insulin delivery system for smart healthcare applications is proposed in \cite{joshi2022iglu,jain2020iglu}.
In the field of smart healthcare, physiological signal monitoring is crucial, and various systems have been developed with the aid of electrocardiography (ECG), photoplethysmography (PPG), and electroencephalography (EEG) monitoring \cite{sharma2021dephnn}\cite{Lin2006-es}\cite{SH_ECG2}. Arrhythmia identification \cite{SH_arrythmia}, heart rate monitoring and chronic heart failure prediction are some applications of ECG monitoring \cite{Seshadri2019}.
The patient's medical history and the outcomes of diagnostic tests are frequently kept on file in the Electronic Health Record (EHR) system of the relevant diagnostic institution or hospital, along with the ECG recordings. The hybrid CVD prediction model can be paired with a cloud or server that extracts traits from the EHR and selects the ones important for CVD prediction. The model can make predictions and select the key qualities on its own. The anticipated outcome could be given to the patient or the relevant clinician, and the conceptual diagram of the framework is shown in Figure \ref{SHC}.

Identifying the nominal sets of characteristics would aid in its precise prediction. Various feature selection methods are reviewed in \cite{frontier_FS_review, ScienceDirect_FS_review}. \cite{9489057} proposed XG Boost model for early prediction of CVD and utilised the RF for feature selection.
An early-stage cardiovascular disease might be difficult to diagnose due to the absence of particular symptoms. Several risk factors, including hypertension, diabetes, and arrhythmias, contribute to the illness. Angiography, echocardiography, and electrocardiography are well-recognized clinical tools for evaluating the risk of CVD. Although angiography is highly precise, it is intrusive and costly. Only cardiologists would be able to interpret echocardiograms and electrocardiography. This constraint has been eliminated by developing autonomous and efficient machine learning algorithms that can forecast the disease based on echocardiogram and electrocardiography characteristics, including demographic and physiological data.
The applicability of machine learning methods, particularly Support Vector Machine (SVM) and boosting algorithms \cite{Krittanawong2020}, for the prediction of CVD appears encouraging. \cite{ghorbani2020deep} demonstrated that the EchoNet deep learning network can accurately analyze echocardiogram pictures to diagnose CVDs. However, evaluating a high number of characteristics at a time is not much required, which leads to large and complex model requirements.
The paper includes 52 characteristics for CVD prediction from four distinct domains: demographic, ECG and laboratory, symptom and examination, and echocardiography, and determines the subset of features that can assist to have accurate CVD estimation.

The following is the paper's structure: 
Section \ref{prior_work} examines a variety of previous publications and strategies for predicting CVDs. Section \ref{reasearch_gap_contribution} describes the research gap and contribution of the paper, while section \ref{data} provides an overview of the data along with a detailed technique. Section \ref{result_discussion} contains findings, commentary, and a comparison of the performance of other models. It also verifies the model with a combined heart disease dataset. In the last section \ref{conclusion}, conclude the article, and outline the future scope of the study.

\section{Prior Works on CVD Prediction} 
\label{prior_work}

\begin{figure}[ht]
\centering
\includegraphics[width=16.5cm]{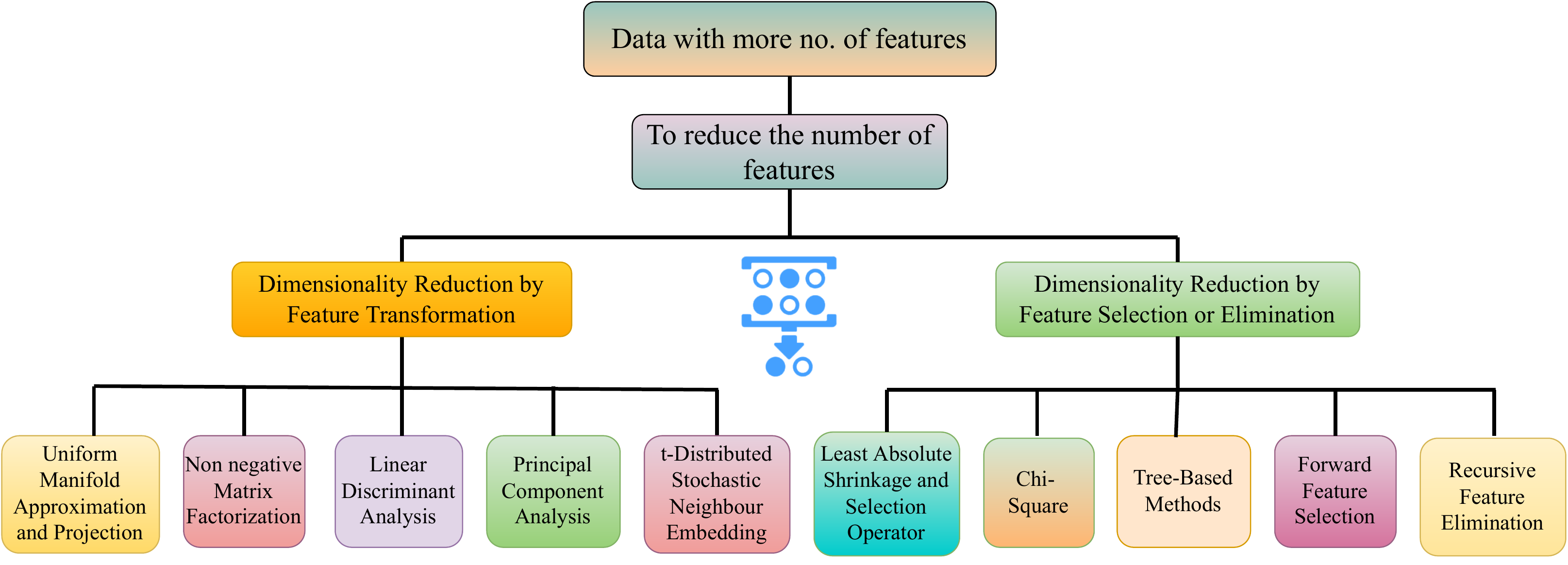}
\caption{Techniques for dimension reduction} 
\label{pf:01}
\end{figure}

Several notable machine-learning models for the classification of cardiovascular disease have been designed in the past few years. Few of them attempted to identify the most important factors for CVD prognosis, while others have worked towards minimising the number of features. Both techniques for improving CVD categorization are illustrated in Figure \ref{pf:01}.

Various feature selection techniques, such as filter-wrapper-based, Infinite feature selection, LASSO, and Ridge, have been utilised to enhance CVD prediction \cite{9802107}\cite{Hasan2021}\cite{Panda_2019}. However, \cite{shilaskar2013feature} utilised forward feature inclusion and backward feature reduction to predict CVD. The related literature review is shown in Table \ref{tab:LR} of the various feature selection for CVD.

\begin{table}[ht]
\centering
\caption{Previous study of Feature Selection Methods and CVD predicting Models} \label{tab:LR}
\begin{tabular}{m{3.8cm}m{3.0cm}m{3.6cm}m{3.2cm}} \toprule
    \textbf{Study} & \textbf{Data set} & \textbf{Feature Selection Methods} & \textbf{Classification Model} \\ \midrule
    Abdellatif et-al, 2022 \cite{9802107} & Statlog and HD clinical records from UCI & Infinite feature  selection (Inf-FS) & Improved weight RF (IWRF) optimized with Baysian optimization \\ \hline
    Hasan et-al, 2021 \cite{Hasan2021} & Kaggle heart disease data set & Filter- wrapper embedded. & RF, SVM, KNN, NB, XGBoost \\ \hline
    Segura et-al,2020 \cite{9281168} & Sleep Heart Health study (SHSS) dataset & PCA and lowest p-value logistic regression & NB, Feed Forward Neural Network, SVM,  RF\\ \hline
    Marbaniang et al,2020 \cite{9342297} & Cardiovascular disease data set from Kaggle & Added two features BP and BMI & KNN Naive Bayes, DT, RF, SVM and LDA \\ \hline
    Panwar et-al,2020 \cite{9180636} & PPG-BP dataset from Figshare & None & Cardio-Net (CNN based Model) \\ \hline
    Panda et-al, 2019 \cite{Panda_2019} & Cleveland Heart Disease Dataset from UCI & LASSO and Ridge & RF, Extra Tree classifier, Gaussian NB, LB \\ \hline
    Shilaskar et-al, 2013 \cite{SHILASKAR20134146} & Frank Ascunion 2010 dataset from UCI. & Forward feature inclusion Back-elimination Forward feature selection & SVM \\ \bottomrule
\end{tabular}
\end{table}

\cite{qian2022cardiovascular} claimed that systolic blood pressure, triglyceride blood glucose index, age, low-density lipoprotein-L/high-density lipoproteins-C, body adiposity index and body mass index are important features for the onset of CVD prediction. Similarly, \cite{Yang2020} designed a CVD risk assessment model using Random Forest(RF) classifier for the eastern Chine population, Whereas \cite{Pasha_2020} proposed improved CVD prediction accuracy using neural networks. \cite{Yazdani2021} proposed an algorithm to calculate the strength score for the features used for CVD prediction. \cite{SWATHY2022109, Taylor_and_Francis21} presented a study of CVD prediction using ML and Deep Learning. \cite{Pal2022-kx} also tried to classify CVD using K-nearest neighbour (K-NN) and multi-layer perceptron (MLP).  \cite{Jiang2021-yp} designed the SVM and LR-based ML model to identify the incident of CVD in the Kazakh Chinese population. Various machine learning algorithms were compared by \cite{CVD_ML_2021} for CVD prediction, and the relation between diabetes and its influences on heart diseases has also been explored.

On chosen fifty MIT-BIH ECG entries, \cite{5643150} presented a real-time CVD classification model capable of alerting the hospital through SMS/MMS or email in the event of an emergency. In contrast, \cite{9491140} provided a comparative examination of the utilisation of different ML models for diverse healthcare applications. The authors proposed a logistic regression model for predicting the threat of cardiovascular disease and diabetes. In addition, they emphasised the significance of the ML approach for predicting serious illnesses related to CVD.
\cite{Muhammad2020} described several ML algorithms for the prognosis of heart illness and developed an ML-based intelligent predictive model for heart disease prediction utilising Cleveland and Hungarian heart disease data sets.
In \cite{Krittanawong2020}, the features of Coronary Computed Tomography Angiography (CCTA) images were also subjected to various Machine Learning (ML) algorithms for CAD classification, including Logistic Regression (LR), Linear Discriminant Analysis (LDA), Decision Tree (DT), Artificial Neural Network (ANN), Support Vector Machine (SVM), and K-Nearest Neighbour (KNN). The authors state that the SVM with a polynomial kernel is the best approach, where the model's parameters were adjusted using the grid search method, and the resulting accuracy was 100\%.
\cite{Krittanawong2020} used a meta-analysis to demonstrate the critical role that machine learning algorithms play in helping clinicians evaluate data and select the best algorithm for the given data set. Additionally, authors have asserted that ML algorithm and Electronic Health Record (EHR) system integration is promising. This can also be employed in smart healthcare architecture to enhance the delivery of high-quality healthcare services \cite{joshi2020secure}.
cardiovascular disease mainly occurs when the heart cannot pump enough blood to the body's organs which is known as Coronary Artery Disease (CAD). When the coronary artery that provides blood to the heart is clogged, the heart's or cardiac muscles' need for oxygen is unmet. 
Stenosis or atherosclerosis are the medical terms for heat-related illnesses.
Artificial intelligence with machine learning models has been widely explored in healthcare, including heart disease prediction. In \cite{aicvd}, the possible application of artificial intelligence to predicting cardiovascular illness was considered.
The mHealth and telemedicine have a great potential to educate clinicians through artificial intelligence techniques and help to have personalised smart healthcare. A review paper that concisely explains the many applications of AI in the topic above is published in \cite{SilviaHindawi}.


\section{Research Gaps and Novel Contribution}
\label{reasearch_gap_contribution}
\subsection{Research Question Addressed in the current Paper}
Still, cardiologists can only diagnose cardiovascular disease after invasive and expensive testing. In the era of smart healthcare and telemedicine, the application of machine learning in CVD diagnosis can be beneficial at an early stage. Despite this, it is still difficult to achieve its clinical performance standards. 
For every machine learning model to be generalizable, two elements are essential. The set of input features comes first, followed by the volume of training data.
Giving the model a tonne of features does not necessarily improve performance \cite{pancholi2021intelligent}. However, it may lead to more complex models and requires large resources. Therefore, the main challenges for Machine Learning techniques in CVD predictions are:\\
(1)  how to choose the appropriate set of features to classify CVD ?\\
(2) which is a suitable machine learning classifier for selected features?

\subsection{iCardo: Proposed CVD Prediction Model}
To address the above challenges, we proposed a hybrid machine learning model called iCardo in the current paper that first selects appropriate sets of features from the database and later utilizes those features for CVD classification. The proposed classification model can also be integrated with a smart healthcare framework. Here, in this proposed study, the optimal model for particular feature sets for cardiovascular prediction is investigated. The original data accounted for characteristics from several fields, including demographic, ECG, symptom and assessment, laboratory, and echo features.
Four feature selection approaches, Least Absolute Shrinkage and Selection Operator (LASSO), Recursive Feature Elimination (RFE), Chi-Square, and Tree-based model, have been used to pick the pertinent subset of features. Later, there are 16 different feature sets obtained through several machine learning classifiers such as Support Vector Machine (SVM), Logistic Regression (LR), Adda Boost, XG-Boost, Naive Bayes, K-Nearest neighbour (KNN), and simple artificial neural network (ANN) and overall 112 models have been assessed. The performance is measured and subsequently compared to discover the best fit for the prediction of CVD. 

\begin{figure}[h]
    \centering
    \includegraphics[width=16cm]{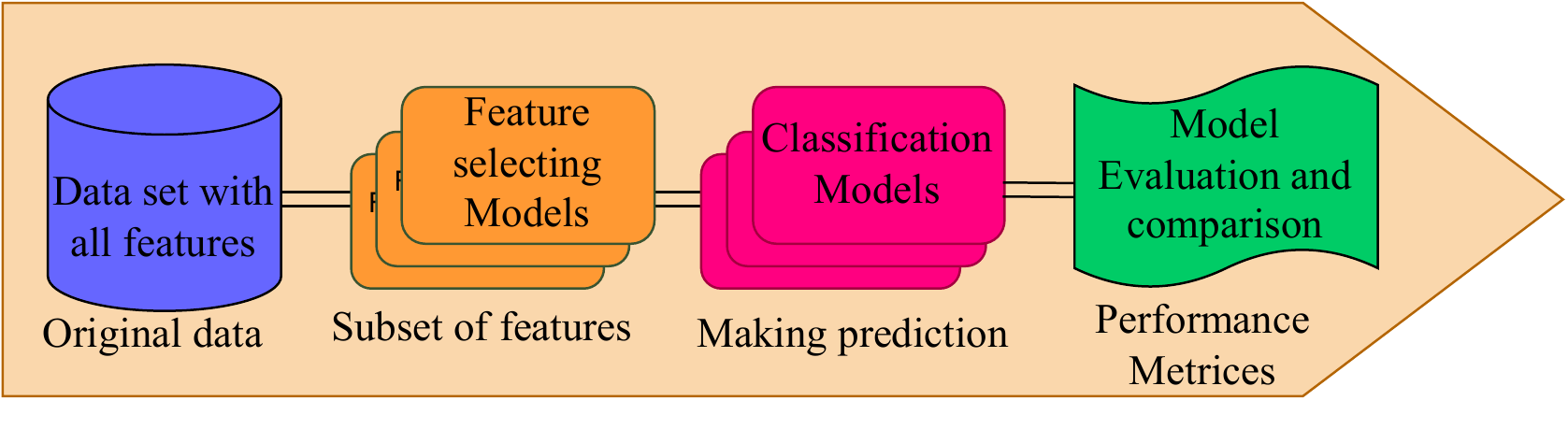}\\
    \caption{Overall process flow of the work}      
    \label{FS-THEME} 
\end{figure}

\begin{figure}[h]
\centering
    \includegraphics[width=11cm]{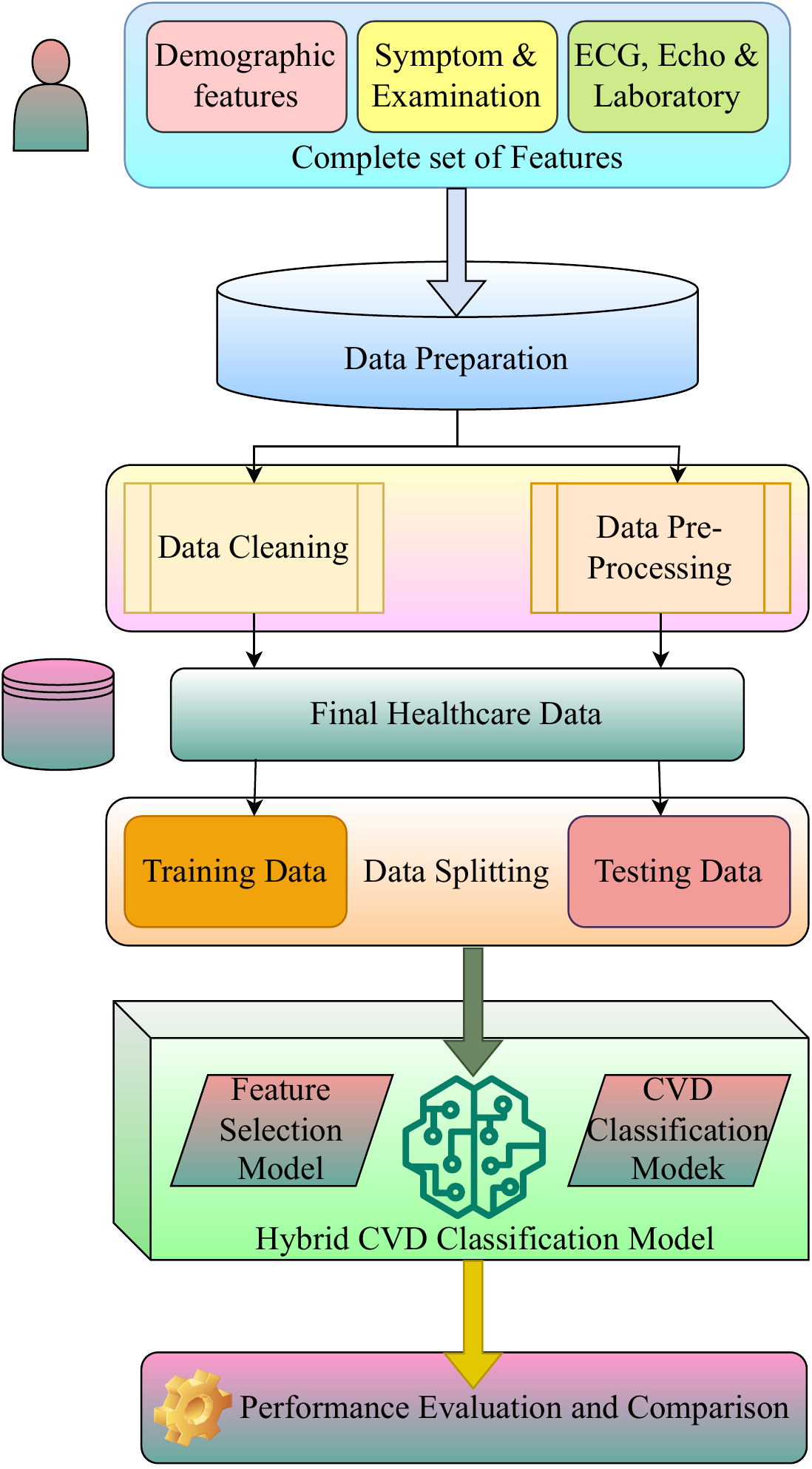}\\
    \caption{Process Flow of iCardo model}
    \label{pf}       
\end{figure}

\subsection{Significance of Proposed Solution}

The significance of the proposed solution are as follow:
\begin{itemize}
    \item Several feature selection models have been trained to select more appropriate features for CVD prediction 
    \item  The classification models are trained with the selected set of features for CVD classification
    \item The performance of the hybrid classification model is presented through comparative analysis
    \item The performance of the proposed model is validated with the help of four publicly available data-sets
    \item The present work would allow to have preventive healthcare mechanisms through a smart healthcare framework
\end{itemize}

\section{CVD Prediction using Hybrid Machine Learning Model}
\label{data}



\subsection{Data set}
\label{sec:2}
Z Alizadeh Sani data set \cite{ARABASADI201719}, \cite{ALIZADEHSANI201352}, and \cite{ALIZADEHSANI2016187} are retrieved from the Machine Learning Repository of UCI and utilised in this study. There are 303 individuals in the dataset, 216 of whom have coronary artery disease. There are 54 properties in this set of information, including ECG, symptom and examination, laboratory and echo, and demographic aspects. Table \ref{tab:1} contains a list of these traits along with their range and category.

\begin{table*}[ht]
\centering
\caption{Features of the dataset}
\label{tab:1}       
\resizebox{\textwidth}{!}
{\begin{tabular}{clcc}  \toprule
    \textbf{Features Name} & \textbf{Range} & \textbf{Feature Type}  \\ \midrule
    f1  & age & 48-120 &	Demographic \\
    f2 & Sex & Female, male &	Demographic \\
    f3  & Diabetes mellitus (DM) & 30-86 & Demographic \\
    f4  & Ex-smoker &	No, yes &	Demographic \\
    f5  & Current smoker &	No, yes &	Demographic \\
    f6  & Hyper tension (HTN) &	No, yes	& Demographic \\
    f7  & Family history (FH) & No, yes &	Demographic \\
    f8  & Body mass index (BMI) (Kg/m2) &	18-41 &	Demographic \\
    f9  &    Dyslipidemia (DLP) &	No, yes &	Demographic \\
    f10 &    Airway Disease &	No, yes	 & Demographic \\
    f11 &    Chronic Renal Failure (CRF) & No, yes &	Demographic \\
    f12 &    Cerebrovascular Accident (CVA)	& No, yes & Demographic \\
    f13 &    Congestive Heart Failure (CHF) &0 No, yes & Demographic \\
    f14 &    Obesity &	Yes, if MBI $>$ 25, No-otherwise &	Demographic \\
    f15 &    Thyroid Disease	& No, yes & Demographic \\
    f16 &    Edema &	No, yes &	Symptom and examination \\
    f17 &    Systolic murmur &	No, yes	& Symptom and examination \\
    f18 &    Typical Chest Pain &	No, yes	& Symptom and examination \\
    f19 &    Atypical & No, yes &	Symptom and examination \\
    f20 &    Weak peripheral pulse & No, yes	& Symptom and examination \\
    f21 &    Exertional Chest Pain (Exertional CP) &	No, yes	& Symptom and examination\\
    f22 &    Nonanginal CP &	No, yes &	Symptom and examination \\
    f23 &    Dyspnea &	No, yes &	Symptom and examination \\
    f24 &    Lung Rales &	No, yes &	Symptom and examination \\
    f25 &    Diastolic murmur &	No, yes	& Symptom and examination \\
    f26 &    low Threshold angina (Low Th Ang) &	No, yes	& Symptom and examination \\
    f27 &    Blood Pressure (BP) (mmHg)	& 90-190 & Symptom and examination \\
    f28 &    Function Class &	1,2,3,4	& Symptom and examination \\
    f29 &    Pulse Rate (PR) (ppm) &	50-110	& Symptom and examination \\
    f30 &    ST Elevation &	No, yes	& ECG \\
    f31 &    Poor R Wave Progression (Poor R Progression) &	No, yes &	ECG \\
    f32 &    T inversion &	No, yes &	ECG \\
    f33 &    Q Wave	& No, yes  &	ECG \\
    f34 & LVH (Left Ventricular Hypertrophy) & No, Yes & ECG \\
    f35 &    ST Depression &	No, yes	& ECG \\
    f36 &   Rhythm & Sin, AF & ECG \\
    f37 &   Lymph (Lymphocyte) (\%) &	7-60	&Laboratory and echo \\
    f38 &    K (Potassium) (mEq/lit)	& 3.0-6.6 &	Laboratory and echo \\
    f39 &    Valvular Heart Disease (VHD) &	Normal, Mild, Moderate, Severe	& Laboratory and echo\\
    f40 &    Blood Urea Nitrogen (BUN) (mg/dl) &	6-52	& Laboratory and echo \\
    f41 &    Creatine (Cr) (mg/dl) &	0.5-2.2 &	Laboratory and echo \\
    f42 &    Low density lipoprotein (LDL) (mg/dl)	& 18-232 & Laboratory and echo \\
    f43 &    Triglyceride (TG) (mg/dl) & 37- 1050 &	Laboratory and echo \\
    f44 &    Erythrocyte Sedimentation rate (ESR) (mm/h) &	1-90 & Laboratory and echo\\
    f45 &    Neutrophil (Neut) (\%)	& 32-89 &	Laboratory and echo \\
    f46 &    High density lipoprotein (HDL) (mg/dl) &	15-111	 & Laboratory and echo \\
    f47 &    Haemoglobin (HB) (g/dl) &	8.9-17.6 & Laboratory and echo \\
    f48 &    Platelet (PLT) (1000/ml) &	25-742 &	Laboratory and echo \\
    f49 &    Fasting Blood Sugar (FBS) (mg/dl) &	62- 400	& Laboratory and echo \\
    f50 &    Sodium (Na) (mEq/lit)	& 128- 156 &	Laboratory and echo \\
    f51 &    Regional Wall Motion Abnormality (Region with RWMA) & 0,1,2,3,4 &	Laboratory and echo \\
    f52 &    Ejection Fraction (EF) &	15,60	& Laboratory and echo \\
    f53 &    White Blood Cell (WBC) (cells/ml) &	3700-18000 & Laboratory and echo \\
    f54 &    Bundle Branch Block(BBB) & No, LBBB, RBBB & Laboratory and echo \\
    f55 &   Weight  & 48 - 120 & Demographic\\
    f56 &   Length  & 140-188 & Demographic\\ \bottomrule
\end{tabular}}
\end{table*}

\subsection{Procedure}
To forecast CVDs, a mixed machine learning approach is suggested. It is divided into two stages: the feature selection stage and the categorization step. 
The classification stage uses seven different classification models, including Logistic Regression (LR), Support Vector Machine (SVM), Adda Boost, XG-Boost, K-Nearest Neighbor (KNN), Naive Bayes, and one simple artificial neural network. The first stage uses four feature selection models that use RFE, Chi-square, LASSO, and tree-based methods.
At the first stage, each feature selection model produces four feature sets with 10, 15, 20, and 25 features. At the first step, a total of 16 feature sets are created, which are mentioned in Table \ref{tab:2}.
Seven categorization methods have subsequently been used for performance evaluation. To choose the most precise model for CVD prediction, the accomplishment of each model is compared. The list of each model's performance metrics is reported in Table \ref{tab:3}.

\begin{table}[ht]
\label{tab:2}
\caption{Features selected in different feature sets}
\resizebox{\linewidth}{!}
{
    \begin{tabular}{ccccccccccccccccc}
    \toprule
     & \multicolumn{4}{c}{\textbf{RFE}} & \multicolumn{4}{c}{\textbf{LASSO}} & \multicolumn{4}{c}{\textbf{Chi-Square}} & \multicolumn{4}{c}{\textbf{Tree-based}} \\ \cline{2-17} 
    \multirow{-2}{*}{\textbf{Features}} & \textbf{1R} & \textbf{2R} & \textbf{3R} & \textbf{4R} & \textbf{1L} & \textbf{2L} & \textbf{3L} & \textbf{4L} & \textbf{1C} & \textbf{2C} & \textbf{3C} & \textbf{4C} & \textbf{1T} & \textbf{2T} & \textbf{3T} & \textbf{4T} \\ \midrule
    f1 &  &  &  &  &\checkmark &\checkmark &\checkmark &\checkmark &  &\checkmark &\checkmark &\checkmark &\checkmark &\checkmark &\checkmark &\checkmark \\ 
    f2 &  &  &\checkmark &\checkmark &  &  &  &  &  &  &  &  &  &  &  &  \\ 
    f3 &\checkmark &\checkmark &\checkmark &\checkmark &\checkmark &\checkmark &\checkmark &\checkmark &\checkmark &\checkmark &\checkmark &\checkmark &  &  &  &  \\ 
    f4 &  &  &  &  &  &  &  &  &  &  &  &  &  &  &  &  \\ 
    f5 &\checkmark &\checkmark &\checkmark &\checkmark &  &  &  &  &  &  &  &\checkmark &  &  &  &  \\ 
    f6 &\checkmark &\checkmark &\checkmark &\checkmark &\checkmark &\checkmark &\checkmark &\checkmark &\checkmark &\checkmark &\checkmark &\checkmark &  &  &  &\checkmark \\ 
    f7 &  &  &  &\checkmark &\checkmark &\checkmark &\checkmark &\checkmark &  &  &  &  &  &  &  &  \\ 
    f8 &  &  &  &  &  &  &  &  &  &  &  &  &\checkmark &\checkmark &\checkmark &\checkmark \\ 
    f9 &  &  &  &  &  &  &  &  &  &  &  &  &  &  &  &  \\ 
    f10 &  &  &  &\checkmark &  &  &  &\checkmark &  &  &\checkmark &\checkmark &  &  &  &  \\ 
    f11 &  &  &  &  &  &  &  &  &  &\checkmark &\checkmark &\checkmark &  &  &  &  \\ 
    f12 &  &  &  &  &  &  &  &  &  &  &  &  &  &  &  &  \\ 
    f13 &\checkmark &\checkmark &\checkmark &\checkmark &  &  &  &  &  &  &  &  &  &  &  &  \\ 
    f14 &  &  &  &  &  &  &  &  &  &  &  &  &  &  &  &  \\ 
    f15 &  &  &  &  &  &  &  &  &  &  &  &  &  &  &  &  \\ 
    f16 &  &  &  &  &  &  &  &  &  &  &  &\checkmark &  &  &  &  \\ 
    f17 &  &  &  &\checkmark &  &  &  &  &  &  &  &  &  &  &  &  \\ 
    f18 &\checkmark &\checkmark &\checkmark & \checkmark &\checkmark &\checkmark &\checkmark &\checkmark &\checkmark &\checkmark &\checkmark &\checkmark &\checkmark &\checkmark &\checkmark &\checkmark \\ 
    f19 &\checkmark &\checkmark &\checkmark &\checkmark &  &\checkmark &\checkmark &\checkmark &\checkmark &\checkmark &\checkmark &\checkmark &\checkmark &\checkmark &\checkmark &\checkmark \\ 
    f20 &  &\checkmark &\checkmark &\checkmark &  &  &  &\checkmark &  &  &\checkmark &\checkmark &  &  &  &  \\
    f21 &  &  &  &  &  &  &  &  &  &  &  &  &  &  &  &  \\ 
    f22 &\checkmark &\checkmark &\checkmark &\checkmark &\checkmark & \checkmark &\checkmark &\checkmark &\checkmark &\checkmark &\checkmark &\checkmark &  &  &  &  \\ 
    f23 &  &  &\checkmark &\checkmark &  &  &  &  &  &\checkmark &\checkmark &\checkmark &  &  &  &  \\ 
    f24 &\checkmark &\checkmark &\checkmark &\checkmark &  &  &  &  &  &  &  &  &  &  &  &  \\ 
    f25 &\checkmark &\checkmark &\checkmark & \checkmark &  &\checkmark &\checkmark &\checkmark &\checkmark &\checkmark &\checkmark &\checkmark &  &  &  &  \\ 
    f26 &  &  &  &  &  &  &  &  &  &  &  &\checkmark &  &  &  &  \\ 
    f27 &  &  &  &  &  &\checkmark &\checkmark &\checkmark &  &  &\checkmark &\checkmark &\checkmark &\checkmark &\checkmark &\checkmark \\ 
    f28 &  &  &  &\checkmark &  &  &\checkmark &\checkmark &  &  &\checkmark &\checkmark &  &  &  &  \\ 
    f29 &  &  &  &  &\checkmark &\checkmark &\checkmark &\checkmark &  &  &  &  &  &  &  &\checkmark \\ 
    f30 &  &  &  &\checkmark &  &  &  &\checkmark &\checkmark &\checkmark &\checkmark &\checkmark &  &  &  &  \\ 
    f31 &  &\checkmark &\checkmark &\checkmark &  &  &\checkmark &\checkmark &  &\checkmark &\checkmark &\checkmark &  &  &  &  \\ 
    f32 &\checkmark &\checkmark &\checkmark &\checkmark &\checkmark & \checkmark &\checkmark &\checkmark &\checkmark &\checkmark &\checkmark &\checkmark &  &  &  &  \\ 
    f33 &  &\checkmark &\checkmark &\checkmark &  &  &\checkmark &\checkmark &\checkmark &\checkmark &\checkmark &\checkmark &  &  &  &  \\ 
    f34 &  &  &  &  &  &  &  &  &  &  &  &  &  &  &  &  \\ 
    f35 &  &  &\checkmark &\checkmark &  &  &  &\checkmark &  &\checkmark &\checkmark &\checkmark &  &  &  &  \\ 
    f36 &  &  &  &  &  &  &  &  &  &  &  &  &  &  &  &  \\ 
    f37 &  &  &  &  &  &  &  &  &  &  &  &  &  &\checkmark &\checkmark &\checkmark \\ 
    f38 &  &  &\checkmark &\checkmark &  &\checkmark &\checkmark &\checkmark &  &  &  &  &  &\checkmark &\checkmark &\checkmark \\ 
    f39 &  &  &  &  &  &  &  &  &  &  &  &  &  &  &  &  \\ 
    f40 &  &  &  &  &  &  &  &  &  &  &  &  &  &\checkmark &\checkmark &\checkmark \\ 
    f41 &  &  &\checkmark &\checkmark &  &  &  &  &  &  &  &  &  &  &  &\checkmark \\ 
    f42 &  &  &  &  &  &  &  &  &  &  &  &  &  &  &\checkmark &\checkmark \\ 
    f43 &  &  &  &  &  &  &\checkmark &\checkmark &  &  &  &  &\checkmark &\checkmark &\checkmark &\checkmark \\ 
    f44 &  &  &  &  &  &  &\checkmark &\checkmark &  &  &  &\checkmark &\checkmark &\checkmark &\checkmark &\checkmark \\ 
    f45 &  &  &  &  &  &  &  &  &  &  &  &  &  &  &\checkmark &\checkmark \\ 
    f46 &  &  &  &  &  &  &  &  &  &  &  &  &  &\checkmark &\checkmark &\checkmark \\ 
    f47 &  &  &  &  &  &  &  &  &  &  &  &  &  &  &\checkmark &\checkmark \\ 
    f48 &  &  &  &  &  &  &  &\checkmark &  &  &  &  &  &\checkmark &\checkmark &\checkmark \\ 
    f49 &  &  &  &  &  &\checkmark &\checkmark &\checkmark &  &  &\checkmark &\checkmark &\checkmark &\checkmark &\checkmark &\checkmark \\ 
    f50 &  &  &  &  &  &  &  &  &  &  &  &  &  &  &  &\checkmark \\ 
    f51 &  &\checkmark &\checkmark &\checkmark &\checkmark &\checkmark &\checkmark &\checkmark &\checkmark &\checkmark &\checkmark &\checkmark &\checkmark &\checkmark &\checkmark &\checkmark \\ 
    f52 &  &  &  &  &\checkmark &\checkmark &\checkmark &\checkmark &  &  &  &\checkmark &\checkmark &\checkmark &\checkmark &\checkmark \\ 
    f53 &  &  &  &  &  &  &  &  &  &  &  &  &  &  &\checkmark &\checkmark \\ 
    f54 &  &\checkmark &\checkmark &\checkmark &  &  &  &  &  &  &  &  &  &  &  &  \\ 
    f55 &  &  &  &  &  &  &  &  &  &  &  &  &  &  &\checkmark &\checkmark \\ 
    f56 &  &  &  &  &  &  &  &  &  &  &  &  &  &  &  &\checkmark \\ \bottomrule
    
    \end{tabular}
}
\end{table}

\subsection{Simulation}
Sklearn is utilised for the complete performance metric evaluation, while Python 3.0 is used for feature extraction, model training, and model testing. Pandas are used to assist in importing and pre-processing data. MinMax scalar is employed to standardise or normalise the data \cite{kokate2021classification}. In addition, Label Encoding is carried out to manage the values, which are alphanumeric and string. Using seaborn and matplot lib, the data and outcomes are visualised through various plots.

\section{Experimental Results}
\label{result_discussion}
This section compares the outcomes of several categorization models using various input characteristics. First, all 56 characteristics of the Z Alizadeh Sani dataset are used with machine learning classifiers, and feature selection approaches. The 10, 15, 20, and 25 features are then extracted using the RFE, LASSO, Chi-Square, and Tree-Based algorithms. Further, each set of features is retrieved using one of  the feature selection methods, and every classifier is trained and evaluated. Table \ref{tab:3} lists the performance indicators for each situation.

\subsection{Feature Selection}
The original dataset comprises 56 characteristics from several categories i.e demographic features, feature-based on symptoms and examinations, feature-based on electrocardiograms, feature-based on lab tests, and feature-based on echocardiography that may be utilised to produce illness outcomes. This dataset  offers a wide variety of features, which makes it better suited for testing feature selection methods. Four feature selection models, given in Table \ref{tab:2}, are used to choose a total of sixteen feature sets.


\begin{sidewaystable}[h]
\centering
\caption{Performance Metrics of different machine learning model on various feature set}
\label{tab:3}
\resizebox{\textwidth}{!}{
\begin{tabular}{llllllllllllllllll}
\toprule
\multirow{2}{*}{\textbf{Models}} &
  \multirow{2}{*}{} &
  \multicolumn{4}{l}{\textbf{10 feature set}} &
  \multicolumn{4}{l}{\textbf{15 feature set}} &
  \multicolumn{4}{l}{\textbf{20 feature set}} &
  \multicolumn{4}{l}{\textbf{25 feature set}} \\
\noalign{\smallskip}\hline\noalign{\smallskip}
 &
   &
  \textbf{Accuracy} &  \textbf{Precision} &  \textbf{Recall}&  \textbf{F1 score} &  \textbf{Accuracy} &  \textbf{Precision} &  \textbf{Recall} &  \textbf{F1 score} &  \textbf{Accuracy} &  \textbf{Precision} &  \textbf{Recall} &  F\textbf{1 score} &  \textbf{Accuracy} &  \textbf{Precision} &  \textbf{Recall} &  \textbf{F1 score} \\ \midrule

\multirow{4}{*}{\textbf{SVM}}         & RFE        & 87.91 & 89    & 64    & 74    & 92.21 & 84    & 84    & 84    & 92.21 & 84    & 84    & 84    & 92.31 & 85    & 88                                & 86    \\ 
                             & Lasso      & 91.2  & 90    & 76    & 83    & 90.11 & 90    & 72    & 80    & 91.2  & 90    & 76    & 83    & 90.11 & 86    & 76    & 81    \\ 
                             & Chi-Square & 89    & 86    & 72    & 78    & 92.31 & 95    & 76    & 84    & 91.21 & 95    & 72    & 82    & 92.31 & 95    & 76    & 84    \\ 
                             & Tree based & 91.21 & 87    & 76    & 83    & 91.21 & 87    & 80    & 83    & 83.51 & 71    & 68    & 69    & 81.31 & 67    & 64    & 65    \\ \hline
\multirow{4}{*}{\textbf{XG-Boost}}    & RFE        & 85.7  & 83    & 60    & 70    & 87.9  & 82    & 72    & 77    & 91.2  & 90    & 76    & 83    & 91.2  & 95    & 72                                & 82    \\ 
                             & Lasso      & 89    & 89    & 68    & 77    & 86.8  & 88    & 60    & 71    & 87.9  & 85    & 68    & 76    & 86.8  & 84    & 64    & 73    \\ 
                             & Chi-Square & 91.2  & 90    & 76    & 83    & 92.3  & 95    & 76    & 84    & 91.2  & 90    & 76    & 83    & 92.3  & 95    & 76    & 84    \\ 
                             & Tree based & 90.1  & 94    & 68    & 79    & 90.1  & 90    & 72    & 80    & 90.1  & 90    & 72    & 80    & 89    & 86    & 72    & 78    \\ \hline
\multirow{4}{*}{\textbf{Adda Boost}}  & RFE &  81.32 &  89 & 85 &  87 &  85.7 &   92.2 &  88.06 &  90 &   81.32 &  90.62 &  84.06 &  87.21 &  83.52 &  92.19 &  85.51
                             &   88.73 \\
                             & Lasso      & 86.61 & 90.62 & 87.87 & 89.23 & 85.7  & 93.75 & 86.95 & 90.22 & 85.71 & 90.62 & 89.23 & 89.92 & 90.11 & 96.87 & 89.85 & 93.33 \\
                             & Chi-Square & 86.81 & 92.18 & 89.39 & 90.76 & 86.81 & 90.62 & 90.62 & 90.62 & 87.91 & 92.18 & 90.77 & 91.47 & 84.61 & 89.06 & 89.06 & 89.06 \\
                             & Tree based & 86.81 & 93.75 & 88.23 & 90    & 85.71 & 92.18 & 88.05 & 90.07 & 82.41 & 89.06 & 86.36 & 87.69 & 84.61 & 96.87 & 83.78 & 89.85 \\ \hline
\multirow{4}{*}{\textbf{Log. Reg.}}   & RFE        & 83.52 & 73    & 70    & 72    & 87.91 & 79    & 81    & 80    & 86.81 & 80    & 74    & 77    & 85.71 & 77    & 74                                & 75    \\ 
                             & Lasso      & 89.01 & 81    & 81    & 81    & 86.81 & 80    & 74    & 77    & 89.01 & 84    & 78    & 81    & 90    & 88    & 78    & 82    \\ 
                             & Chi-Square & 87.91 & 79    & 81    & 80    & 89.01 & 84    & 78    & 81    & 87    & 81    & 78    & 79    & 86.81 & 80    & 74    & 77    \\ 
                             & Tree based & 84.62 & 76    & 70    & 73    & 85.71 & 77    & 74    & 75    & 72.53 & 56    & 37    & 44    & 75.82 & 63    & 44    & 52    \\ \hline
\multirow{4}{*}{\textbf{KNN}}         & RFE        & 82.42 & 70    & 70    & 70    & 84.62 & 78    & 67    & 72    & 83.52 & 75    & 67    & 71    & 81.32 & 71    & 63                                & 67    \\ 
                             & Lasso      & 76.92 & 71    & 37    & 49    & 74.73 & 67    & 30    & 41    & 72.53 & 58    & 26    & 36    & 72.53 & 67    & 15    & 24    \\ 
                             & Chi-Square & 85.71 & 79    & 70    & 75    & 78.02 & 67    & 52    & 58    & 74.73 & 62    & 37    & 47    & 70.33 & 50    & 22    & 31    \\ 
                             & Tree based & 70.33 & 50    & 19    & 27    & 69.23 & 43    & 11    & 18    & 67.03 & 20    & 4     & 6     & 67.03 & 20    & 2     & 6     \\ \hline
\multirow{4}{*}{\textbf{Naive Bayes}} & RFE        & 82.41 & 72    & 63    & 68    & 52.75 & 39    & 100   & 56    & 52.75 & 39    & 100   & 56    & 57.14 & 41    & 96                                & 57    \\ 
                             & Lasso      & 87.01 & 81    & 78    & 79    & 85.71 & 75    & 78    & 76    & 72.53 & 52    & 100   & 68    & 62.44 & 44    & 96    & 60    \\ 
                             & Chi-Square & 51.25 & 38    & 100   & 55    & 53.85 & 39    & 100   & 56    & 61.54 & 43    & 96    & 60    & 61.54 & 43    & 96    & 60    \\ 
                             & Tree based & 87.91 & 75    & 89    & 81    & 84.62 & 72    & 78    & 75    & 84.62 & 72    & 78    & 75    & 81.32 & 68    & 70    & 69    \\ \hline
\multirow{4}{*}{\textbf{ANN}}         & RFE        & 82.41 & 61    & 77    & 68    & 82.41 & 65    & 65    & 65    & 81.32 & 79    & 54    & 64    & 80.22 & 62    & 81                                & 70    \\ 
                             & Lasso      & 87.91 & 76    & 73    & 74    & 79.12 & 60    & 96    & 74    & 81.31 & 68    & 54    & 60    & 76.92 & 56    & 72    & 63    \\ 
                             & Chi-Square & 87.91 & 82    & 79    & 81    & 86.81 & 81    & 76    & 79    & 82.41 & 80    & 71    & 75    & 83.51 & 86    & 48    & 62    \\ 
                             & Tree based & NA    & NA    & NA    & 73.62 & 73.62 & 37    & 48    & 30    & 80.22 & 59    & 83    & 69    & 71.42 & 0     & 0     & 0    \\ \hline
\end{tabular}}
\end{sidewaystable}

\subsection{Performance Comparison}

\begin{figure}[ht]
\centering
\subfigure[]{
  \includegraphics[scale =0.85]{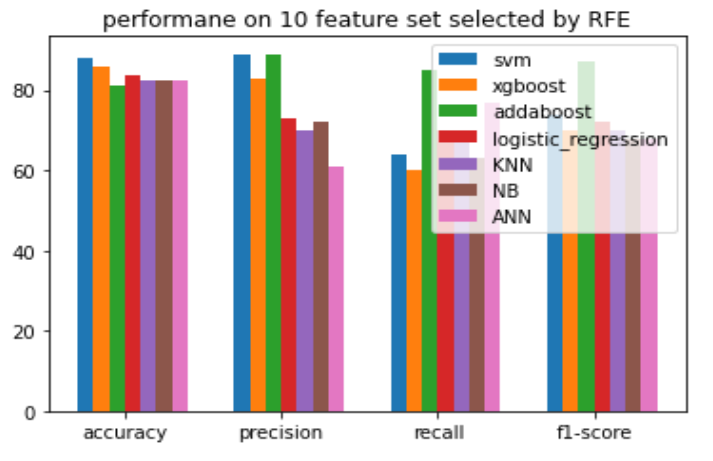}
  \label{10RFE}
 }
 \subfigure[]{
  \includegraphics[scale =0.85] {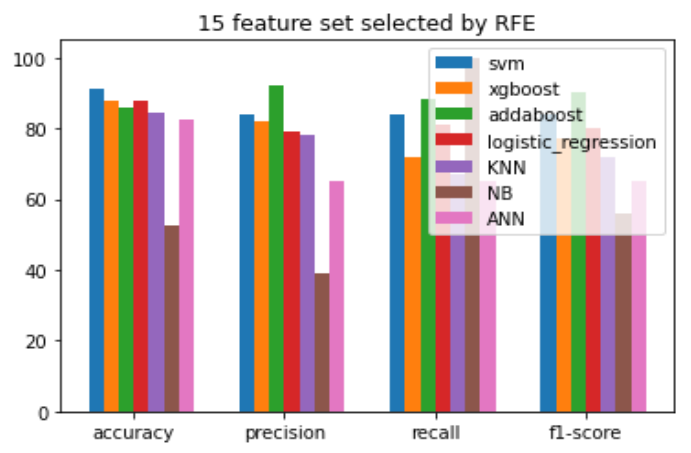}
  \label{15RFE}
 }
\subfigure[]{
  \includegraphics[scale =0.85] {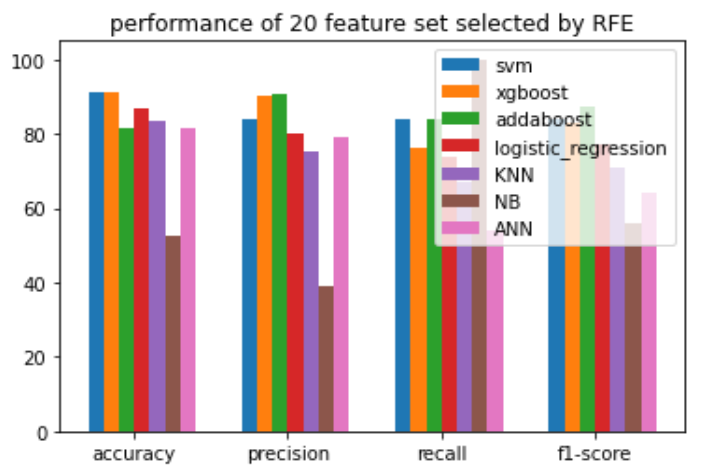}
  \label{20RFE}
 }
\subfigure[]
{
  \includegraphics[scale =0.85]{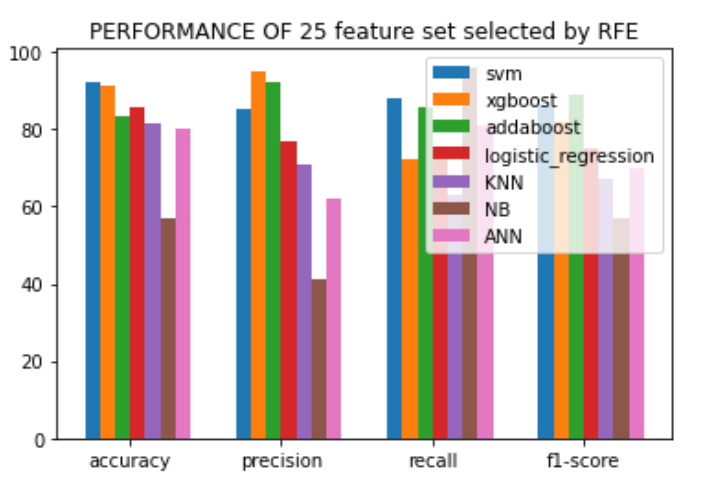}
  \label{25RFE}
}

\caption{Performance of different ML models on the feature sets selected by RFE feature selection model (a) 10 features, (b) 15 features, (c) 20 features, (d) 25 features}
\label{RFE}
\end{figure}

\begin{figure}[ht]
\centering
\subfigure[]{
  \includegraphics[scale =0.85] {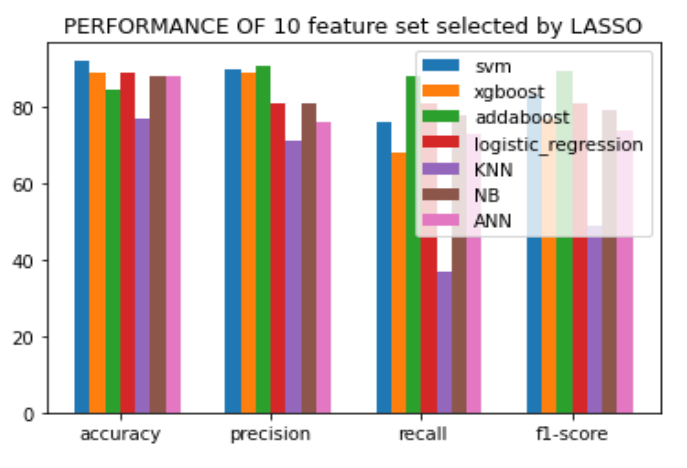}
  \label{10LASSO}
 }
 \subfigure[]{
  \includegraphics[scale =0.85] {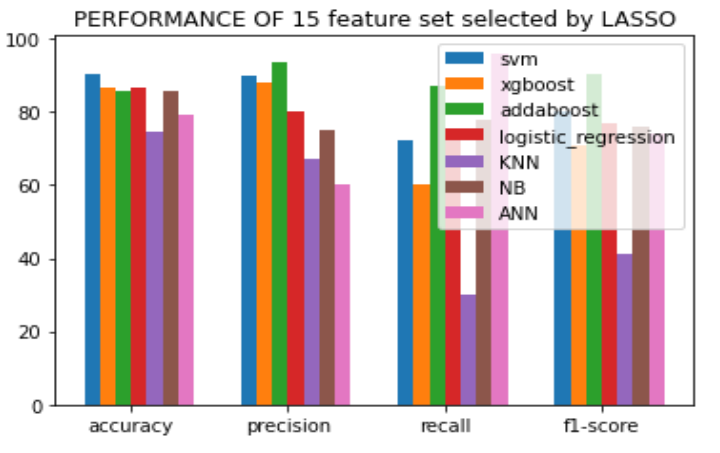}
  \label{15LASSO}
 }
\subfigure[]{
  \includegraphics[scale =0.85] {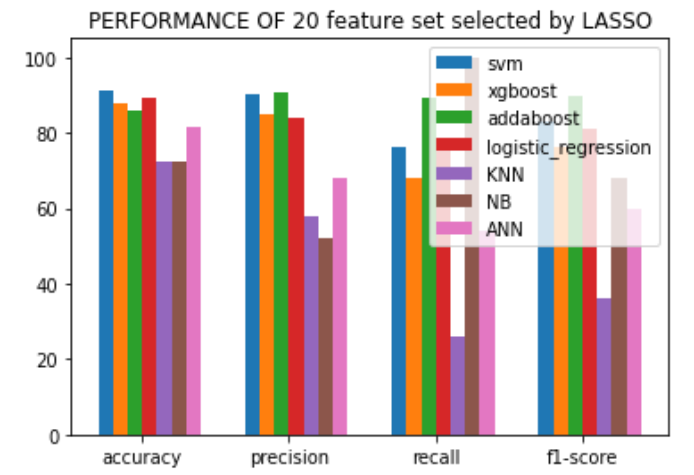}
  \label{20LASSO}
 }
\subfigure[]{
  \includegraphics[scale =0.85] {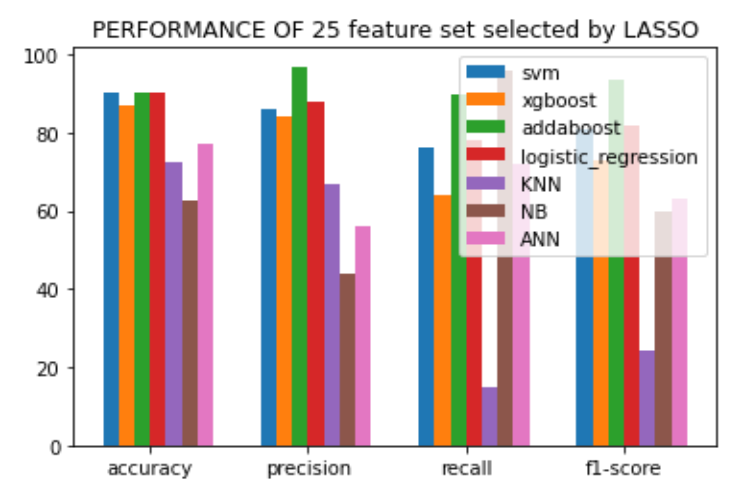}
  \label{25LASSO}
 }

\caption{Performance of different ML models on the feature sets selected by Chi-square feature selection model (a) 10 features, (b) 15 features, (c) 20 features, (d) 25 features}
\label{LASSO}
\end{figure}

\begin{figure}[ht]
\centering
\subfigure[]{
  \includegraphics[scale =0.85] {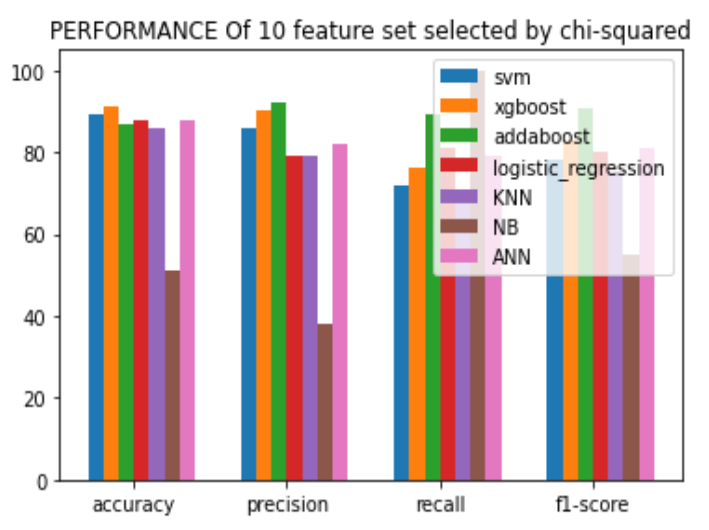}
  \label{10CS}
 }
 \subfigure[]{
  \includegraphics[scale =0.85] {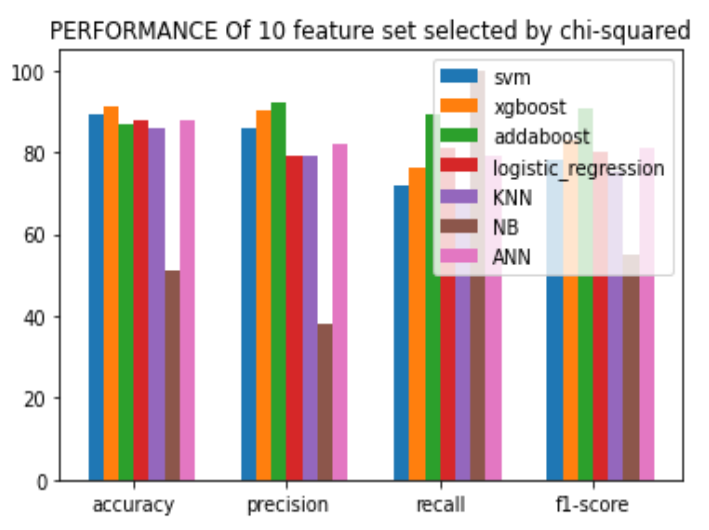}
  \label{15CS}
 }
\subfigure[]{
  \includegraphics[scale =0.85] {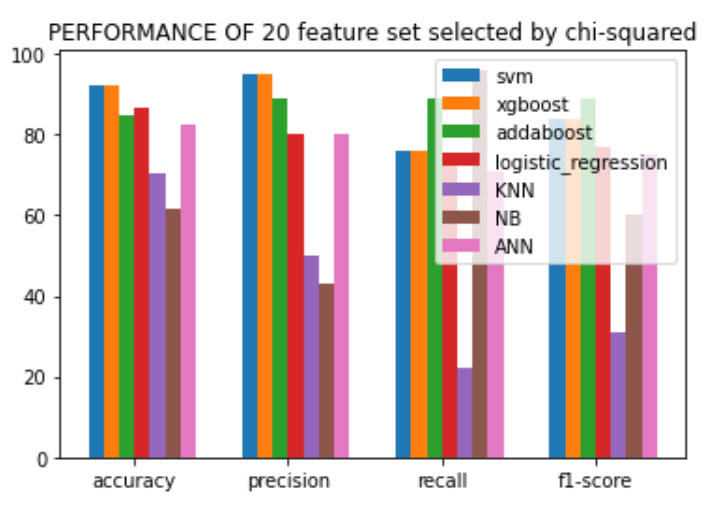}
  \label{20CS}
 }
\subfigure[]{
  \includegraphics[scale =0.85] {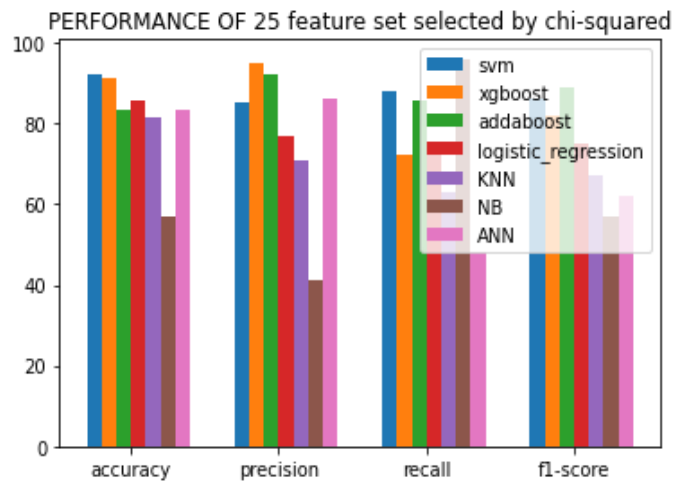}
  \label{25CS}
 }

\caption{Performance of different ML models on the feature sets selected by LASSO feature selection model (a) 10 features, (b) 15 features, (c) 20 features, (d) 25 features}
\label{CS}
\end{figure}

\begin{figure}[ht]
\centering
\subfigure[]{
  \includegraphics[scale =0.85] {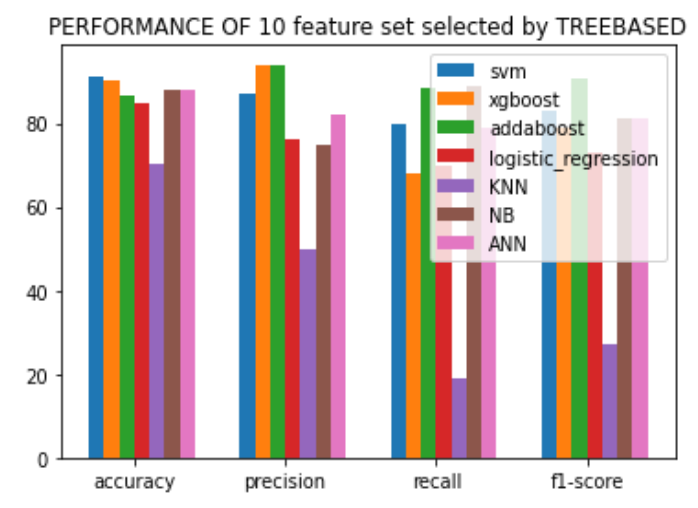}
  \label{10TB}
 }
 \subfigure[]{
  \includegraphics[scale =0.85] {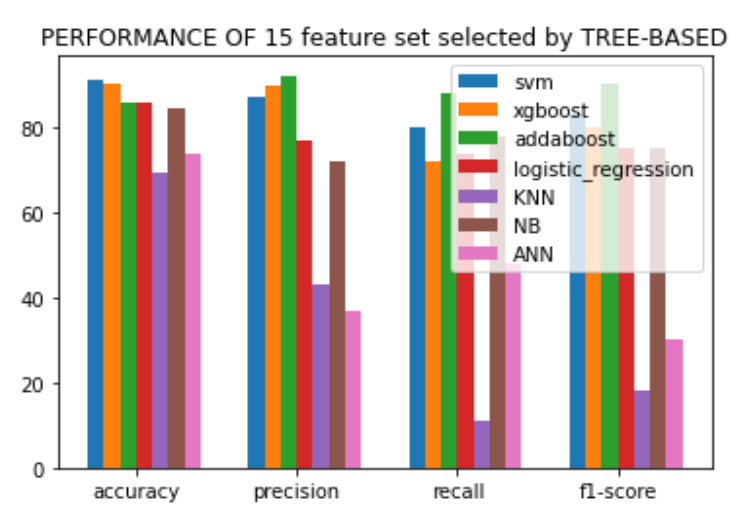}
  \label{15TB}
 }
\subfigure[]{
  \includegraphics[scale =0.85] {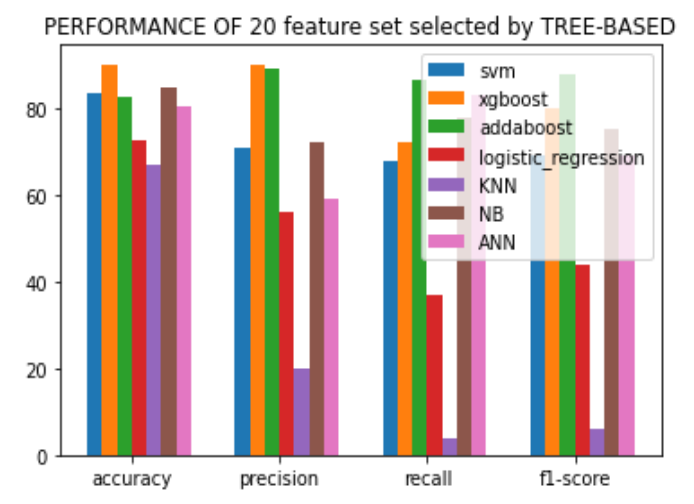}
  \label{20TB}
 }
\subfigure[]{
  \includegraphics[scale =0.85] {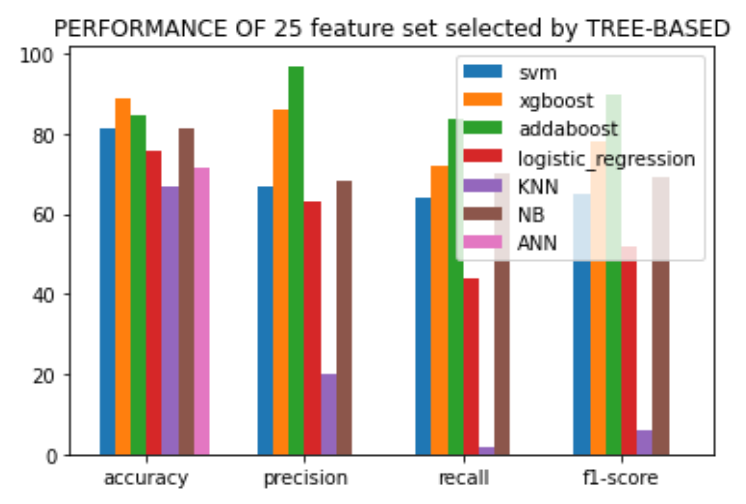}
  \label{25TB}
 }
\caption{Performance of different ML models on the feature sets selected by Tree-Based feature selection model (a) 10 features, (b) 15 features, (c) 20 features, (d) 25 features}
\label{TreeBased}
\end{figure}

A comparison has been performed regarding the overall accuracy of the models. 
The 15 feature sets are chosen using the Chi-square model and the 25 feature sets produced using RFE and the Chi-square model have the best SVM accuracy, which is 92.31 \%. SVM, however, performs better in terms of complexity and resource usage \cite{joshi2022fpga}. For the same 15 characteristics of the Chi-square model, the accuracy, recall, and F1-score are 95 \%, 76 \%, and 81 \%, respectively. Precision is crucial in addition to overall accuracy since it indicates what percentage of a positive forecast is accurate \cite{sinha2022predicting}. With the 25 features obtained using the tree-based feature selection model, Adda Boost has the highest precision, which is 96.87 \%, however, adaptive boosting is a more sophisticated technique and requires more resources. With the Chi-square model-derived set of 15, 20, and 25 features, SVM likewise produces high accuracy, which is 95\%. Figure \ref{RFE}, \ref{LASSO}, \ref{CS} and \ref{TreeBased} display the performance of various machine learning models on feature sets chosen using various feature selection algorithms.


The ratio of accurately anticipated positives to all actual positives is provided by the recall, true positive rate, sensitivity, or hit rate. For 10 feature sets chosen by the Chi-square model, 15 feature sets chosen by the RFE and Chi-square model, and 20 feature sets chosen by the RFE and LASSO model, the value of recall for Naive Bayes is greatest, being 100\% for each of these feature sets. The model's approach is based on the fact that each character is independent of the others, which is not quite true in this case. 
For instance, ageing affects hypertension (HTN). Similar to this, diabetes mellitus (DM) may be inherited, develop from obesity, or both, and may contribute to congestive heart failure (CHF). The SVM classifier can be suggested for predicting cardiovascular disease based on the aforementioned justifications.

\begin{figure}[ht]
\centering
    \includegraphics[width=7cm]{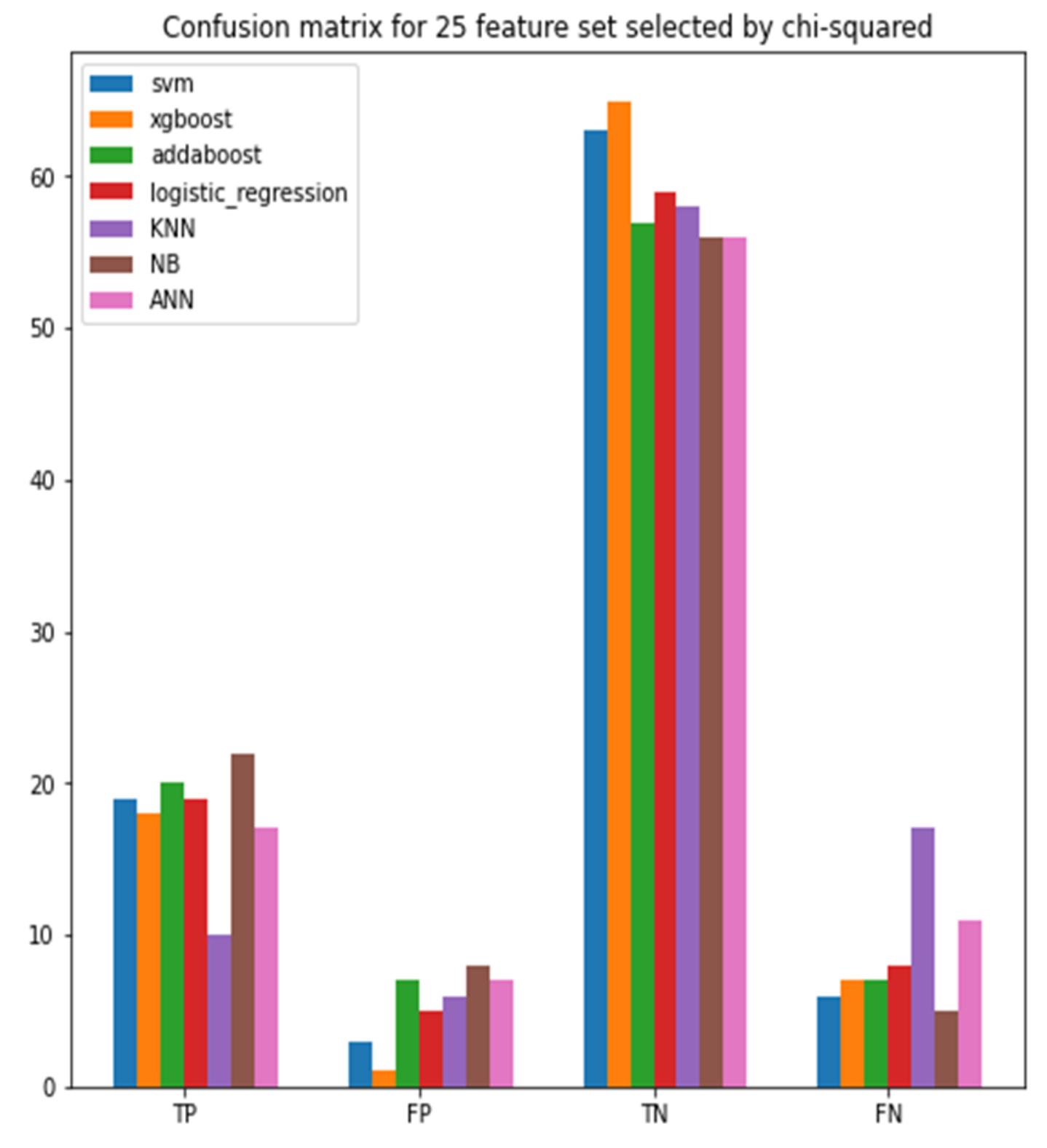}\\
    \caption{Bar graph of attributes of confusion matrix}
    \label{CF4C}       
\end{figure}

The confusion matrix is one of the finest ways to evaluate a classification model's effectiveness \cite{pancholi2021novel}. Figure \ref{CF4C} displays the bar graph of the confusion matrix for several machine learning models. SVM classifiers with RFE feature selection models often provide the highest level of accuracy for CVD classification.

\subsection{Validation}
Four data sets,  Cleveland, Hungarian, Switzerland, and Long Beach, VA have been combined to validate the performance of the suggested model. 
There are thirteen features in the dataset, including serum cholesterol level (chol), age, sex, maximum heart rate (thalach), chest pain (cp), exercise-induced angina (exang), resting blood pressure (trestbps), fasting blood sugar (fbs), resting ECG result (restecg), the slope of the peak exercise ST segment (slope), ST depression caused by exercise relative to rest (oldpeak), number of major vessels (0-3) coloured by flouroscopy (ca) (thal). The combined data is run through an SVM model, which produced an accuracy of 78\%; however, when the data is run through the suggested model, the accuracy is increased to 83\%. Table \ref{tab:4} compares several techniques based on CVD classification accuracy. 


\begin{table}[ht]
\caption{Comparison with other Literature Work}
\label{tab:4}
\resizebox{\textwidth}{!}
{
    \begin{tabular}{llll} \toprule
    Previous Work & Data & Technique  & Accuracy \\
    \midrule
    
    Babi et-al, 2017 \cite{babivc2017predictive} & Cleveland, Hungarian, Switzerland and & Neural Networks & 89.93 \\
                   & Long Beach VA, and South &     &        \\
                   &  Africa Heart Disease data  & SVM & 73.70 \\
                   & Z-Alizadeh Sani Dataset     & SVM & 86.67 \\
       \hline            
    Haq et-al, 2018 \cite{Hybrid2018}  & Cleveland heart disease & Hybrid Intelligent System & 89 \& 88 \\
    \hline
    Mohan et-al, 2019 \cite{effective} & Cleveland dataset & \begin{tabular}[c]{@{}l@{}}Hybrid Random Forest with\\ a Linear Model (HRFLM).\end{tabular} & 88.70 \\
    \hline
    Ali et-al, 2019 \cite{8684835} & Cleveland heart disease database & Stacked SVM based expert system & 91.11 \\
    \hline
    
    Hashi et-ql, 2020 \cite{developing} &  Cleveland Heart Disease dataset & Hyper-parameter Tuned SVM & 89.23 \\
    \hline
    Spencer et-al, 2020  \cite{spencer2020exploring} & Combination of four data set & BayesNet & 85.00 \\
                           & (Cleveland, Long-Beach-VA, Hungarian &     \\
                           &  and Switzerland dataset) &    \\
    
                        & data set 2016 &  Framework LR and SVM   &   \\
                        \hline
    Dissanayake et-al, 2021 \cite{dissanayake2021comparative} & Cleveland Heart Disease dataset & Decision Tree & 88.52 \\
    
    \hline
    Pal et-al, 2022 \cite{Pal2022-kx} & CVD data from UCI & K-NN, MLP & 73.77, 82.47 \\
    \hline
    \textbf{Proposed} & Z-Alizadeh Sani data & SVM with RFE & \textbf{92.31} \\ 
                      & Combined heart disease dataset & SVM (without feature selection) & 78.00 \\
                      & Cleveland, Switzerland, Hungarian, & SVM with RFE & \textbf{82.10} \\
                      &  and Long-beach-VA                &               &            \\ \bottomrule
            
    \end{tabular}
}
\end{table}

\section{Conclusion}
\label{conclusion}
The early identification of heart disease can help for preventing premature death through effective treatment. The analysis of several characteristics for CVD prediction using ML models is presented in this paper. With the best accuracy of 92.31\%, SVM surpasses all other ML models for all feature sets, and this is accomplished by using 15 and 25 feature sets chosen by chi-square models and 25 feature sets chosen by RFE models, respectively. The chi-square model is discovered to be the best for feature selection. Overall, it can be claimed that SVM with RFE is the most effective method for classifying CVDs.
In future research, we will try to evaluate the proposed hybrid model with a bigger data set with a vast range of features. We will also consider the integration of the proposed model with Electronic Health Records through Internet-of-Medical Things (IoMT) platform. It would allow to have remote monitoring of critical condition of CVD patient for smart healthcare management.

\clearpage
\bibliographystyle{unsrt}
\bibliography{Trefrence}

\section*{About the Authors'}


{\hbox{\fbox{\includegraphics[width=2cm]{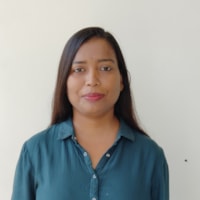}}}}%
{\begin{minipage}{0.95\textwidth}  
		\textbf{Nidhi Sinha}(Member, IEEE) is a research scholar in Electronics and Communication Department at MNIT Jaipur, Rajasthan, and doing her research in field of Early detection of Cardiovascular Failure using Machine Learning and AI. She is also working as Data Scientist at Konverge.AI, Pune (remotely). She has experience in the Machine Learning and Artificial Intelligence project deployment as well as in the field of solar cell design and has published research papers in International Journals and Conferences.\par
\end{minipage} }

{\hbox{\fbox{\includegraphics[width=2cm]{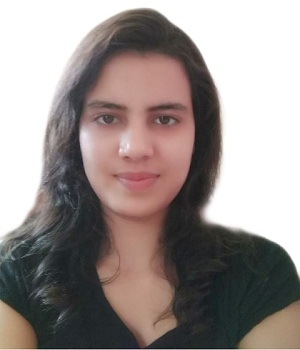}}}}%
{\begin{minipage}{0.95\textwidth}  
		\textbf{Teena Jangid} has completed her post-graduation in VLSI-Design from MNIT Jaipur, Rajasthan. She has done research in the field of Early detection of Cardiovascular Failure using Machine learning and AI. She is currently working as Design Verification Engineer at MeyvnSystems Pvt Ltd.\par
\end{minipage} }



{\hbox{\fbox{\includegraphics[width=2cm]{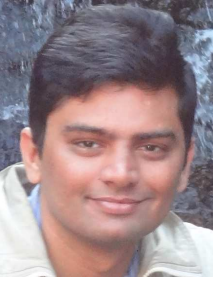}}}}%
{\begin{minipage}{0.95\textwidth}  
		\textbf{Dr. Amit M. Joshi} (Senior Member, IEEE) completed his M.Tech and PhD from NIT, Surat, in 2009 and 2015, respectively. He is currently an Assistant Professor at Malaviya National Institute of Technology, Jaipur (MNIT Jaipur) since July 2013. His area of specialization is Biomedical signal processing, Smart healthcare, VLSI DSP Systems, and embedded system design. He is a senior member of IEEE, IETE and a member of IEEE. He has published 100+ research articles in excellent peer-reviewed international journals/conferences and six book chapters. He has a total of 1000+ google scholar citations, an i10 index of 33, and an H-index of 17. He served as a reviewer of technical journals such as IEEE Transactions IEEE Access, Springer, and Elsevier and as a Technical Programme Committee member for IEEE conferences (iSES, ICCE, ISVLSI, VDAT). He also received the honour of a UGC Travel fellowship, the SERB DST Travel grant award, and a CSIR fellowship and also attended well-known IEEE Conferences TENCON-16, TENCON-17, ISCAS-18, MENACOMM-19, etc. across the world. He has also served as Mentor for IEEE Engineering in Medicine and Biology Society student mentorship program 2021. He has supervised 4 PhD thesis and 26 M.Tech projects, and 17 B.Tech projects in Biomedical Signal Processing, VLSI/Embedded Systems. \par
\end{minipage} }

{\hbox{\fbox{\includegraphics[width=2cm]{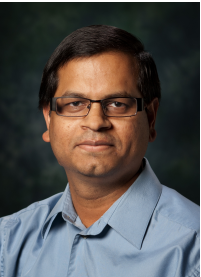}}}}%
{\begin{minipage}{0.95\textwidth}
		\textbf{Saraju P. Mohanty} (Senior Member, IEEE) received a bachelor's degree (Honors) in electrical engineering from the Orissa University of Agriculture and Technology, Bhubaneswar, in 1995, the master's degree in Systems Science and Automation from the Indian Institute of Science, Bengaluru, in 1999, and the Ph.D. degree in Computer Science and Engineering from the University of South Florida, Tampa, in 2003. He is a Professor with the University of North Texas. His research is in "Smart Electronic Systems" which has been funded by National Science Foundations (NSF), Semiconductor Research Corporation (SRC), US Air Force, IUSSTF, and Mission Innovation. He has authored 450 research articles, 4 books, and invented 9 granted/pending patents. His Google Scholar h-index is 47 and i10-index is 203 with 10300 citations. He is a recipient of 14 best paper awards, Fulbright Specialist Award in 2020, IEEE Consumer Electronics Society Outstanding Service Award in 2020, the IEEE-CS-TCVLSI Distinguished Leadership Award in 2018, and the PROSE Award for Best Textbook in Physical Sciences and Mathematics category in 2016. He has delivered 15 keynotes and served on 13 panels at various International Conferences. He has been the Editor-in-Chief of the IEEE Consumer Electronics Magazine during 2016-2021 and currently serves on the editorial board of six journals/transactions. \par
\end{minipage} }

\end{document}